\documentclass[12pt,fleqn]{article}
\usepackage{xiiiemc}
\usepackage{natbib}
\usepackage{fancyhdr}
\usepackage{color}
\usepackage{wallpaper} 
\usepackage{titlesec}   %% Define space between paragraph e section
\usepackage{float} 	%% Use to fix Figure or Table: ex: \begin{table}[H]
\usepackage[usenames,dvipsnames]{xcolor}
\usepackage{hyperref} 	%% Use to fix Figure or Table: ex: \begin{table}[H]
\hypersetup{
	pdfcreator={LaTeX with abnTeX2},
	pdfkeywords={abnt}{latex}{abntex}{USPSC}{trabalho acadêmico}, 
	colorlinks=true,       		% false: boxed links; true: colored links
	linkcolor=blue,          	% color of internal links
	citecolor=blue,        		% color of links to bibliography
	filecolor=magenta,      		% color of file links
	urlcolor=blue,
	allbordercolors=black,
	bookmarksdepth=4
}
\usepackage{tocloft}
\titleformat{\section}
  {\normalfont\bfseries}{\thesection.}{0.5em}{}
 
\renewcommand\thesection{\arabic{section}}

\let\OLDthebibliography\thebibliography
\renewcommand\thebibliography[1]{\OLDthebibliography{#1} \setlength{\parskip}{0pt}\setlength{\itemsep}{0pt plus 0.3ex}}

\title{EGALITARIAN ASPECTS OF SCALE-FREE NETWORKS}

\author
    {\rm \begin{tabular}{l} 
    \textbf{Renato Fabbri}$^{1}$ - {\textnormal renato.fabbri@gmail.com}\\%
    \textbf{Marília M. Pisani}$^{2}$ - {\textnormal marilia.m.pisani@gmail.com}\\
    {\fontsize{11}{0}\selectfont $^{1}$University of São Paulo, Institute of Mathematical and Computer Sciences - São Carlos, SP, Brazil}\vspace*{-0.05cm} \\
    {\fontsize{11}{0}\selectfont $^{2}$Federal University of ABC, Centre for Natural Sciences and Humanities - São Paulo, SP, Brazil}\vspace*{-0.05cm}\\
  \end{tabular}}
\fancypagestyle{firspagetstyle}
{
	\lhead{}
	\fancyhead[C]{%
		\includegraphics[width=0.9\linewidth]{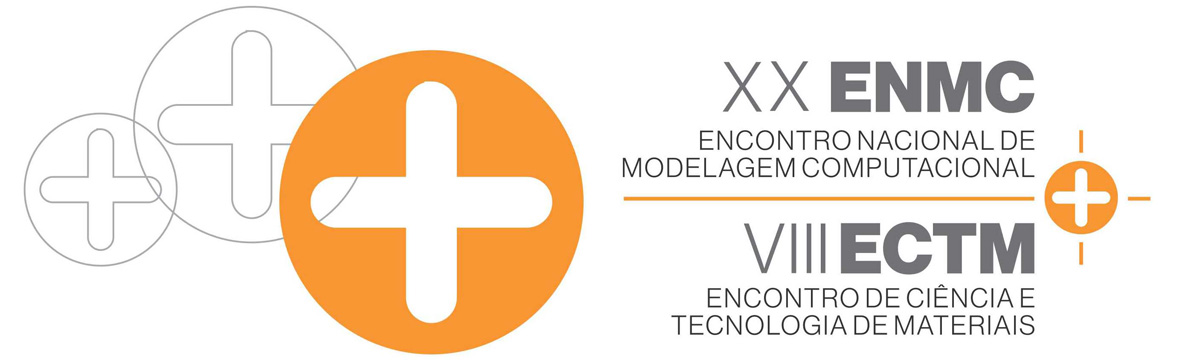}\\%
		{\scriptsize \fontfamily{phv}\fontseries{b}\selectfont \color[rgb]{0.45,0.45,0.45}
		16 a 19 de Outubro de 2017\\
		Instituto Politécnico - Universidade do Estado de Rio de Janeiro\\
		Nova Friburgo - RJ\\
	    }
	}
	\renewcommand{\headrulewidth}{0.0pt}
	\fancyfoot[C]{\footnotesize \parbox{15cm} {\centering  \fontsize{7.5}{0}\selectfont \it Anais do XX ENMC – Encontro Nacional de Modelagem Computacional e VIII ECTM – Encontro de Ciências e Tecnologia de Materiais,  Nova Friburgo, RJ – 16 a 19 Outubro 2017}} % \ttfamil
	\rhead{}
}

\begin{document}
\maketitle

\thispagestyle{firspagetstyle}

\fancyhead[L]{\footnotesize{\fontsize{7.5}{0}\selectfont \it XX ENMC e VIII ECTM\\
	16 a 19 de Outubro de 2017\\
	Instituto Politécnico Universidade do Estado do Rio de Janeiro – Nova Friburgo - RJ\\}}
\renewcommand{\headrulewidth}{0.0pt}
\fancyfoot[C]{\footnotesize \parbox{15cm} {\centering  \fontsize{7.5}{0}\selectfont \it Anais do XX ENMC – Encontro Nacional de Modelagem Computacional e VIII ECTM – Encontro de Ciências e Tecnologia de Materiais,  Nova Friburgo, RJ – 16 a 19 Outubro 2017}} % \ttfamil
\rhead{}

\begin{abstract}
Scale-free networks are frequently described as the zenith of inequality and sometimes even pin-pointed
as a natural cause of concentrations, including accumulation of resources in human society.
Although coherent with theory and empirical data,
there are at least three aspects of scale-free networks that are egalitarian.
Using the social network metaphor (which is easier to grasp):
1) the presence of each agent (vertex or component) in diverse networks,
while each agent has the same amount of resources (e.g. time) for engaging (establishing links or edges) with other individual agents,
ideas or objects;
2) the constant change in the concentration of resources (in this case, links) of each agent;
3) the uniform distribution of resources with respect to the amount of resources per agent (the more resources per agent, the fewer the agents).
We also consider the importance of vertices that are not hubs and overstatements on the importance of the hubs.
The conclusions suggest the urge of research to better examine and model the relevance of the vertices
that are not hubs (i.e. peripheral and intermediary) and
verify the need to better weight the emphasis current literature
places in the association of inequality to the scale-free networks (and other phenomena in which power laws are found).
\end{abstract}


\pagestyle{fancy}

\section{INTRODUCTION}
\cite{price1}
reported that citations networks had a heavy-tailed distribution following a power law.
He also described a ``cumulative advantage'' that explains the power law~\citep{price2}.
This same networks with a power-law distribution of connectivity is nowadays called ``scale-free''
and the same explanation for the distribution is called ``preferential attachment'',
as the result of a rediscovery of the property by~\cite{barabasi1}.
This power-law rediscovery, and the rediscovery of small-world networks by~\cite{small},
is usually considered the birth of the complex networks field,
in the turn of the millennium.
In the last decade and a half,
most of the articles in physics report advances in complex networks.

Another interesting fact about the meaning of the field
is related to the essence of these structures.
Components (be them human agents or not), in diverse contexts, often
exhibit a power-law distribution of activity.
If their activity yields links,
the power-law distribution of connectivity is one of the byproducts of activity.
Indeed, connectivity and activity present high correlation in many of such scenarios~\citep{fabbri1,fabbri2,fabbri3}.

The complex network literature stresses concentration and often revolves around the hub. 
More accurately, it glamorizes the most connected nodes as
``both the strength and the weakness of scale-free networks''~\citep{wikipedia,networks},
and places them as the most important vertexes.
In this work, we argue that this is naive for a number of reasons.
Therefore, this document is countercurrent with respect to available literature:
not only the focus here is to observe scale-free networks egalitarian aspects,
but also to dilute the hubs boast.
No academic writing was found by the authors to expose this simple and pertinent content.

Section~\ref{sec:can} presents a glimpse at standard theory related to power laws.
Section~\ref{sec:three} states the aspects of scale-free networks that are egalitarian.
Section~\ref{sec:delusion} considers also the hubs, intermediary and peripheral vertices
and a probable unbalance current literature has in considering the importance of these sectors.
Section~\ref{sec:con} is dedicated to conclusions and future work.

\section{CANONICAL BACKGROUND}\label{sec:can}
Networks with a power-law distribution of connectivity (degree or strength\footnote{Degree of a vertex
is the number of links connected to it.
Strength of a vertex is the sum of the weight of the links connected to it.})
are called \emph{scale-free}.
In other words, be $p(k)$ the probability that an arbitrary vertex has degree $k$,
than, for a scale-free network, one can assume:
\begin{equation}
p(k) \sim k^{-\gamma}
\end{equation}
\noindent where $\gamma > 1$ is constant and typically $\gamma \in [2,3]$.
This same distribution is called, under certain conditions, e.g. the Pareto distribution or the Zipf's law.
In fact, they all report that nature is performing a distribution of resources in
a way that can be modeled by power laws.
The next sections exposes this content very briefly.

% LER AO MENOS: \url{http://www.nature.com/nature/journal/v435/n7039/full/nature03459.html} e \url{https://www.pik-potsdam.de/members/kurths/publikationen/2010/pnas-18803.full.pdf}.

\subsection{Zipf's law}
This law is unique among the ones mentioned in this article because it
relates the frequency of an element to its rank in an ordering where the
most frequent comes first.
More specifically, a word with rank $r$ has the frequency (or probability of occurring):
\begin{equation}\label{eq:pl}
p(r) \sim r^{-\gamma}
\end{equation}
\noindent where $\gamma \approx 1$.
If the definition given above is followed strictly, the law entails
that:

\begin{equation}
	p(r) = \frac{r^{-\gamma}}{\sum_{n=1}^N n^{-\gamma}}
\end{equation}
\noindent where N is the vocabulary size.
Mandelbrot proposed a shifted version of this law that
follows more closely the distributions found in real languages:

\begin{equation}
	p(r) \sim (r+s)^{-\gamma}
\end{equation}
\noindent where $s\approx 2.7$.

There are many non-trivial considerations to be made
about the Zipf's law but, overall, it seems to hold universally
in all languages, including those that are extinct and untranslated~\citep{zipf}.

\subsection{Pareto distribution}
The Pareto distribution is defined in many ways throughout the literature.
\cite{paretoWikipedia}, for example, defines the distribution through
its survival function and presents 4 variants and a generalized form
with an extra parameter.
In its most basic form, the Pareto distribution can be regarded as
having the probability density function:
\begin{equation}
	p(k) = \frac{\gamma k_m^\gamma}{k^{\gamma+1}}
\end{equation}
\noindent where $k_m$ is the minimum value for $k$ to yield $p(k)>0$.
This distribution is used to describe many observable phenomena,
including in social, financial and geophysical theories.

\subsection{Human perception: Steven's law}
The relation between the magnitude of a physical stimulus and
its perceived intensity is often given by the Steven's law:
\begin{equation}
	S(I) = kI^\gamma
\end{equation}
\noindent where $S(I)$ is the perceived intensity (or strength)
of intensity $I$ of stimulus type that entails $\gamma$ and
$k$ is a proportionality constant that accounts for the arbitrary
units used.

\subsection{Many other laws}
Notice that this last law (Steven's) is not a probability law as in the other cases above.
In fact, many other phenomena are modeled as power laws, especially natural phenomena,
such as Newton's laws of gravitation and inertia. 
This calls for a consideration of the expressive capabilities of power laws and
for an honest inquiry of what is so special about these functions that
can account for its ubiquity~\citep{fabbri3}.

\subsection{Scale-free networks}
Now back to the main subject of this article.
A scale-free network is one where the distribution of connectivity
follows the Equation~\ref{eq:pl}.
These networks also present other non-trivial properties,
such as the 'small-world-ness', high clusterization and
abundant motifs~\citep{networks}.
Most interesting for this discussion, these networks are not isolated
entities, are not static and the power-law distribution is directly related to
an uniform distribution as will be exposed in the next section.

\section{THREE EGALITARIAN ASPECTS}\label{sec:three}
This is the most important section of this article.
We strived to keep everything as simple as possible to allow
both this document to be reasonable to a large audience
and for the reader to promptly grasp the contents.
The kernel of our subject is: given the ubiquity of power-laws,
and that they arise from an uniform distribution,
and that, in the context of social networks, these structures
are not isolated nor static,
what should be the emphasis given to the association of scale-free
networks to inequality?
In the same context: does it explain the ubiquity of inequality
but states, at the same time, paradoxically, that equality is also prevalent?
We should examine each of the egalitarian aspects of scale-free networks:
\begin{itemize}
	\item according to the exposition in~\cite{fabbri3}, 
		a power-law is entailed by a uniform distribution.
		The same quantity $C$ of resources
		is distributed uniformly across all concentrations of resources:
		$\forall k : p_U(k)=C \Rightarrow p(k) = C.k^{-\gamma}$
		where $\gamma$ is the dimensionality of the resource.
                In other words, the uniform distribution of a $\gamma$-dimensional
		resource result is a power law with coefficient $\gamma$.
		The more resources is allocated per component (e.g. per agent), the less numbered
		are the components.
	\item The networks are not static!
		Some of them might be (e.g. those yield by bone cavities)
		but social networks, which are at the center of our discussion, are not.
		Social networks are in constant change and, in fact, stable hubs were only observed
		in very small networks~\citep{barabasi}.
		Although the overall distribution is somewhat stable (or invariant) in the power-law outline,
		the participants move constantly between the periphery to intermediary and hubs sectors~\citep{fabbri1,fabbri2,versinus}.
		Thus, the scale-free (social) networks are egalitarian in the sense that
		the participants present only transient concentrations of resources.
	\item The networks are not isolated!
		This has deep consequences for social networks.
		For example, you, the reader, are a hub in some nucleus of your own family.
		Some friend of yours is a peripheral in your family because he/she only knows you.
		In the same way, everyone is a hub, a peripheral and an intermediary\footnote{A
		sound method for deriving the hubs, intermediary and peripheral sectors,
		by comparing the real network against the Erdös-Rényi model, is described
		in~\cite{fabbri1} and~\cite{fabbri2}.} at the same time,
		but in a (vast) number of different networks.
\end{itemize}

\section{EXALTATION OF HUBS AND DELUSIONS OF GRANDEUR}\label{sec:delusion}
The hubs, defined as the nodes with the greatest number of links in a scale-free network,
are reported in many studies as being the most relevant components.
In fact, selective attacks to the hubs (generally modeled as removing them)
cause the networks to loose many of the properties that make scale-free networks
appealing: the low average-distance between nodes, the fast transmission of information
and other sorts of transportation~\citep{networks}.
Nevertheless, aging of hubs is a recognized phenomenon,
i.e. the descent of the most connected hubs is observed
and even expected~\citep{aging}.
Also, diseases spread rapidly by means of the hubs 
and there are many other important negative contributions
of the hubs and important positive contributions of the
other vertices.
In social networks terms, often used for such discussions:
\begin{itemize}
    \item Hubs are but a few vertexes, which are replaced constantly.
    \item Hubs present the most trivial behavior:
	    they interact as much as possible, in every situation, with every other agent.
		That is a prerequisite for they to be hubs.
    \item Hubs tend to present corrupt behavior,
	    as they apply huge amounts of time and energy (resources)
		on the network where it is a hub and frequently depend on such activity for basic provision.
		This might also be seen as a tendency of the hub to have a committed attitude,
		but given that hubs are also ubiquitous in non-professional networks,
		this might be understood as corrupt behavior by the non-warned.
    \item World input to the network is done mostly by peripheral and intermediary agents,
	    as they do not apply all their energy on the network.
    \item Authorities are often intermediaries or less active hubs,
	    especially in cases where they deliver quality, not quantity.
		This can be noticed by the fact that authorities usually exploit their authority
		in order to achieve their goals by means of the dedication of other agents.
	\item Network structure is yield by intermediary vertices~\citep{fabbri2},
	    as they are the only with non-trivial behavior:
		peripheral vertexes interact only a few times; hubs interact everywhere possible.
\end{itemize}

In summary, the constant exaltation of the hubs in current
complex networks literature seems to be related to the
fact that the field is still incipient.
A psychologist might also add that it is related to
grandiose delusions which affect a great fraction of
our society and thus leaks into the scientific tradition~\citep{grandiose}.
A social scientist might add that there is a constant
pressure performed by the more powerful and rich groups
to legitimate themselves and the inequality as a natural tendency~\citep{socialB,marcuse}.
An anthropologist might mention societies where the inequality 
is symbolized in rituals as something to be avoided
and extreme cases in which the existence of leaders is denied~\citep{clastres}.
It is also appropriate to remember the notes of~\cite{heisenberg}
about nature, physics and human social power.
Anyway, there is an evident consonance between the
constant ennoblement of the hubs, the stress on the association
of scale-free networks and inequality,
and our observation that these associations should be
considered with parsimony.

\section{CONCLUSIONS AND FURTHER WORK}\label{sec:con}
The observations made in this brief article have direct consequences for
the purposes of the mathematical and agent-based models of complex networks.
How can our models reflect the diversity input by peripherals
and the greater authority held by the intermediary sector or the less connected hubs?
Considering our current scientific literature:
should one build a survey of the
excesses about inequality and exaltation of hubs in recent complex networks books, article and software?
More fundamentally, are the issues reported in this article sound
and relevant?
Given that the complex networks field is recent and that it is as driven
by social constructs and conveniences as any knowledge field,
we need to choose a very lucid path
in 1) asking more questions than stating fresh conclusions as facts;
2) establishing a deeper dialog with the traditions that have studied
these phenomena for centuries in advance, such as social scientists and philosophers.

Further work should involve attempts to answer the questions and follow the recommendations above,
further confirm or refute the arguments in this documents, and:
\begin{itemize}
	\item develop models to reflect the fundamental roles performed by nodes beyond hubs (i.e.
		peripherals and intermediary).
	\item Deepen the model exposed in~\cite{fabbri3} and submit it to peer review processes in order
		to understand the extent of its validity.
	\item Study the dynamics of individual agents in social networks with respect to the concentration of links,
		as suggested in~\cite{fabbri1,fabbri2,versinus}.
	\item Study the integration of the networks, maybe by using the database in~\cite{losd} which
		contains a number of networks with shared participants.
	\item Deepen the absorption of the literature and the partnership with academics which produce
		knowledge within the complex networks field, both to better understand the convictions
		and to spread our concerns.
	\item Consider an antithesis: the formalism of institutionalized posts
		might impose both a less dynamic variation of participant roles
		and a less extreme concentration of resources.
\end{itemize}
% Are there other egalitarian aspects which we are not taking into account?

\subsection*{\textit{Acknowledgements}}
The authors thank CNPq for the funding received while researching the topic of this article,
the researchers of IFSC/USP and ICMC/USP for the recurrent collaboration in every situation
where we needed directions for investigation.

% ------------------------------------------------------------------------

% ------------------------------------------------------------------------

%For papers written in Portuguese or Spanish.

%\begin{center}
%  TITLE IN ENGLISH
%\end{center}

%\def\abstractname{Abstract}%

%\begin{abstract}
%   Abstract in english
%\end{abstract}

%\keywords{\em{Keywords in english}}

\end{document}